\documentclass{aa}
\let\linenumbers\relax
\nolinenumbers
\usepackage{graphicx}
\usepackage{xcolor}
\usepackage{txfonts}
\usepackage{comment}
\usepackage[colorlinks=true,
            citecolor=blue,
            linkcolor=blue,
            urlcolor=blue]{hyperref}
\begin{document} 

 \title{Critical Eddington luminosity and photon cooling of compact spherical stars}
   \titlerunning{ Eddington luminosity of compact spherical stars}
   \author{Tisha Desai
          \inst{1},
           Ji{\v r}i Hor\'{a}k
           \inst{2},
           W{\l}odek Klu\'zniak
          \inst{1},
          \and
         Sukalpa Kundu
          \inst{3,1}
          }
         \institute{
         Nicolaus Copernicus Astronomical Center of the Polish Academy of Sciences, Bartycka 18, 00-716 Warsaw, Poland
         \email{tishadesai.work@gmail.com}
         \email{wlodek@camk.edu.pl}
         \and 
         Astronomical Institute, Academy of Sciences, Bo\v{c}n\'{i} II 141 31, Prague 4, Czech Republic
         \email{horak@astro.cas.cz}
         \and
         SGMK Nicolaus Copernicus Superior School, The College of Astronomy and Natural Sciences, Nowogrodzka 47A, 00-697
         Warsaw, Poland
         \email{skundu@camk.edu.pl}         
         }

   \date{Received September 30, 2026}

  \abstract
   {Relativistic corrections to the luminosity have been used for neutron stars of radius greater than $3MG/c^2$ in studies of X-ray bursts. However, there is no fundamental limit that excludes the existence of more compact massive objects with a physical surface, fluid or solid.}
   {We aim to discuss critical "Eddington" luminosity for spherical stars that have a surface within the sphere of unstable circular photon orbits (photon sphere).}
   {We use standard analytical ray-tracing formulae in the Schwarzschild metric.}
  {We find that for stars within the photon sphere, the Eddington condition for balance between radiative force and gravitational attraction is fundamentally altered by photons returning to the surface. The equilibrium is non-local and requires stationarity on timescales of photon fall-back. 
  As a result, spontaneous symmetry breaking is possible, leading to non-uniformity of emission across the photospheric surface.}
  {For a fixed intrinsic luminosity, the photons returning to the surface decrease the cooling rate of stars inside their photon sphere.
  At the same time, the returning photons increase the value of the critical luminosity that would balance gravity. One and the same redshifted formula applies to the critical luminosity, as seen by a distant observer, for any radius of the star. In the X-ray burst literature, for stars with radius less than 3 gravitational radii a different formula has long been accepted that missed the essential physics of the photons emitted in directions outside the escape cone.
  }
   {}

   \keywords{Gravitation --
                Relativistic processes --
                Radiation: dynamics
               }
   \authorrunning{T. Desai et al.}

   \maketitle

%________________________________________________________________

%%%%%%%%%%%%%%%%%%%%%%%%%%%%%%%%%%%%%%%%%%%%%%%%%%%%%%%%%%%%%%%%%%%%%%%%
\section{Introduction}
%%%%%%%%%%%%%%%%%%%%%%%%%%%%%%%%%%%%%%%%%%%%%%%%%%%%%%%%%%%%%%%%%%%%%%%%
At the end stage of stellar evolution, lower-mass stars form white dwarfs, while the more massive ones end up as a neutron star or a black hole \citep{Heger+03}. General relativistic effects become significant in the case of neutron stars, where the typical radius lies around a few times the Schwarzschild radius. A bound on the maximum possible compactness of a neutron star arises from the {Schwarzschild limit for stars made of incompressible matter,  $R>2.25M$, extended to isotropic fluid with no density inversion by}
\cite{Buchdahl59}. A more stringent bound of $R>2.83M$ is obtained via the causality limit that dictates that the sound speed may not exceed the speed of light \citep{LP16}. A recent analysis of a variety of equations of state compatible with {nuclear theory and perturbative quantum chromodynamics} pushed the compactness to $R>3M$ \citep{RE26}. However, one can always consider exotic compact objects that violate the assumptions behind these constraints, such as gravastars of radius only infinitesimally larger than the Schwarzschild radius \citep{MM02}, or a subclass of cloaked naked singularities where the source of gravity is surrounded by a fluid body in hydrostatic equilibrium with a photosphere that lies below the photon circular orbits \citep{VK23}, and similar solutions \citep[e.g.,][]{Cardoso+19}. 

Theoretical constraints notwithstanding,  we would like to have direct observational limits for the radius to mass relation for compact objects endowed with a material surface. If their radius is smaller than that of the unstable circular orbit their emission signatures are expected to differ from standard neutron stars \citep{Nemiroff+93}. Pulse profiles from the polar hot spots on the surface of neutron stars have been used to obtain information about the compactness of the neutron star \citep{Miller+19}.

Another standard tool to estimate the radius of compact stars involves comparing the observed luminosity in flat-top X-ray bursts to the expected critical ("Eddington") luminosity \citep{vP78,E87,Ozel06} . For putative stars of radius smaller than three gravitational radii \citet{Lewin1993} suggested that for photons emitted below the photon sphere (the imaginary sphere on which unstable circular photon orbits lie) the escape cone of solid angle less than $2\pi$ would reduce the observed critical luminosity of X-ray bursts, since not all of the emitted photons escape to infinity in that case. It turns out that including the physics of the photons returning to the surface changes that result. We revisit the issue and relate the observed luminosity to the emitted one for a variety of scenarios.

%%%%%%%%%%%%%%%%%%%%%%%%%%%%%%%%%%%%%%%%%%%%%%%%%%%%%%%%%%%%%%%%%%%%%%%%
\section{Emission from spherical stars}
%%%%%%%%%%%%%%%%%%%%%%%%%%%%%%%%%%%%%%%%%%%%%%%%%%%%%%%%%%%%%%%%%%%%%%%%
We assume the gravitational field of the star to be described by the Schwarzschild metric
\begin{equation}
ds^2 = -H^2(r)\ dt^2 + \frac{1}{H^2(r)} \ dr^2 + r^2\  d\theta^2 + r^2\sin^2\theta \ d\phi^2,
\label{eq:ds2}
\end{equation}
where, $H(r)=\sqrt{1-2M/r}$, and $M$ is the gravitational radius related to the stellar mass, $M_*$, through $M=M_*(G/c^2)$. 

In our calculations we will be considering a spherically symmetric photosphere ("surface of the star") at radius $R$, and isotropic emission.
Outside the photosphere we only consider geodesic motion of photons, neglecting scattering in any medium (e.g., an optically thin corona), if present. We will also briefly discuss possible modifications to our results when these assumptions do not hold.

%%%%%%%%%%%%%%%%%%%%%%%%%%%%%%%%%%%%%%%%%%%%%%%%%%%%%%%%%%%%%%%%%%%%%%%%
\subsection{Angle subtended by the star and the escape cone}
%%%%%%%%%%%%%%%%%%%%%%%%%%%%%%%%%%%%%%%%%%%%%%%%%%%%%%%%%%%%%%%%%%%%%%%%
Assuming  isotropic emission of radiation, the observed flux at a distance $r$ is given by
\begin{align}
F(r)&=\int I(r)\cos(\theta)d\Omega \nonumber \\
    &=I(r)\int_0^{2\pi}d\phi\int_0^{\alpha(r)}
\cos\theta\sin\theta\,d\theta
=\pi I(r)\sin^2\alpha(r)
\label{eq:Lalpha}
\end{align}
\citep{RL79}, where $\alpha(r)$ is the half-angle subtended at $r$ by the star (from the center to the edge of the stellar image on the sky). The luminosity of the star is obtained through integrating the flux over the surface area of a sphere of radius $r$,
\begin{equation}
    L(r) =4\pi r^2F(r).
    \label{lum}
\end{equation}
On the surface of the star at $r=R$, by setting $\alpha=\pi/2$ we get 
\begin{align}
    F(R)&=\pi I(R) \nonumber  \\
    L(R)&=4\pi^2R^2I(R).
\label{eq:LFR}
\end{align}

The angle $\tilde\alpha$ at which a ray arrives at the observer, relative to a line pointing at the center of the star, is related to the angle of emission (at the surface) relative to the zenith, $\tilde\delta(R)$, 
by the equation \citep{AEL90}
\begin{equation}
    \sin\tilde\alpha(r) = \frac{R}{r} \frac{H(r)}{H(R)} \sin\tilde\delta(R).
    \label{eq:sa}
\end{equation}
The half-angle $\alpha(r)$ subtended by the star is the largest value of the angle $\tilde\alpha(r)$.

In the case of ultracompact stars with radius smaller than that of unstable circular photon orbits (the photon sphere at $r=3M\equiv R_{\rm ph}$  in the Schwarzschild metric), 

only photons within the so called escape cone of vertex half-angle $\delta$ are able 

to reach the observer at $r>R_{\rm ph}$. Photons outside this escape cone fall back onto the photosphere.
Thus, only photons with  $\tilde\delta<\delta$ escape to infinity, and photons with $\tilde\delta=\delta$ graze the photon sphere \citep{MTW73}, and hence,
\begin{equation}
    \sin\alpha(r) = \frac{R}{r} \frac{H(r)}{H(R)} \sin\delta(R) , \ \ \ \mathrm{when}\  R< 3M,
    \label{eq:sab}
\end{equation}
with
\begin{equation}
    \sin\delta = \frac{R_{\rm ph}}{R} \frac{H(R)}{H(R_{\rm ph})} =3\sqrt{3}\frac{M}{R}\left(1-\frac{2M}{R}\right)^{1/2} , \ \ \ \mathrm{when}\  R< 3M,
    \label{eq:delta}
\end{equation}
following from Eq.~(\ref{eq:sa}) with $\tilde\alpha(R_{\rm ph})=\pi/2$, which yields
\begin{equation}
    \sin\delta(R) = \frac{3\sqrt{3}M}{R}\left(1-\frac{2M}{R}\right)^{1/2}, \ \ \ \mathrm{when}\  R< 3M.
    \label{eq:sb}
\end{equation}
Thus, in this case ($ R< 3M$) the angle subtended by the star is the angle subtended by the photon sphere \citep{Nemiroff+93}.

If  there are are no unstable circular photon orbits  outside the
photosphere, we take  $\delta(R)=\pi/2$ to compute $\alpha$,  since in that case all of the outgoing emitted photons are able to reach the observer, including those that graze the stellar surface \citep{S66}:
\begin{equation}
    \sin\alpha(r) = \frac{R}{r} \left(\frac{1-2M/r}{1-2M/R}\right)^{1/2}, \ \ \ \mathrm{when}\  R\geq 3M.
    \label{eq:saa}
\end{equation}

%+++++++++++++++++++++++++++++++++++++++++++++++++++++++++++++++++
\begin{figure}
    \centering
    \includegraphics[width=\columnwidth]{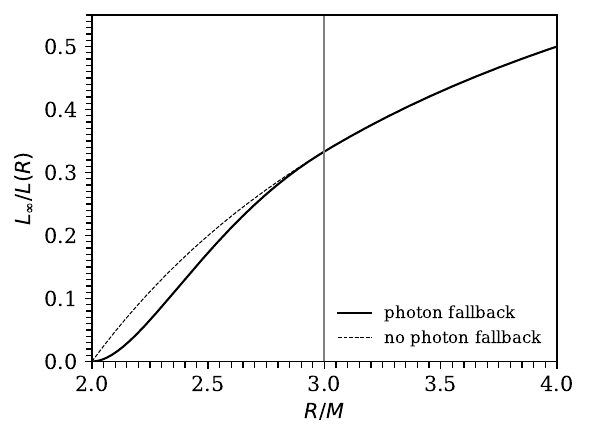}
    \caption{The ratio of the observed luminosity to the surface luminosity is shown as a function of stellar radius. The ratio tends to zero at the Schwarzschild radius $R=2M$, and gradually increases as the stellar radius becomes larger and the gravitational redshift $(1-2M/R)$ becomes smaller, tending to the Newtonian limit of unity at $R\rightarrow\infty$. However, when the surface of the star is below the photon sphere (gray vertical line at $R_{\rm ph}=3M$),  the observed luminosity experiences an additional suppression. The gray dashed line shows the ratio obtained when the effect of the escape cone is ignored.}
    \label{fig:LR_Linf}
\end{figure}
%+++++++++++++++++++++++++++++++++++++++++++++++++++++++++++++++++
%%%%%%%%%%%%%%%%%%%%%%%%%%%%%%%%%%%%%%%%%%%%%%%%%%%%%%%%%%%%%%%%%%%%%%%%
\subsection{Observed luminosity}
%%%%%%%%%%%%%%%%%%%%%%%%%%%%%%%%%%%%%%%%%%%%%%%%%%%%%%%%%%%%%%%%%%%%%%%%
Using the invariance of $I/\nu^4$ one obtains the redshifted intensity $I(r)=I(R)[H(R)/H(r)]^4$.
When there is no unstable circular photon orbit outside the photosphere, Eq.~(\ref{eq:saa}) gives the subtended angle, 
and the luminosity at $r$ is obtained via  the flux given by Eqs.~(\ref{eq:Lalpha}), (\ref{lum}):

\begin{align}
    L(r) &= 4\pi^2 r^2 I(r) \sin^2\alpha(r)= 4\pi^2 R^2 I(R) H^2(R)/H^2(r) \nonumber\\
         &= L(R)H^2(R)/H^2(r).
\end{align}
At $r\to\infty$, this reduces to the usual redshift  formula 
\begin{equation}
    L_\infty=L(R)\,(1-2M/R), \ \ \ \mathrm{when}\  R\geq 3M.
    \label{eq:LoutCO}
\end{equation}

For photospheres within the photon sphere, the angle subtended by the star is obtained using Eq. (\ref{eq:sb}), and hence the luminosity at infinity turns out to be reduced by $\sin^2\delta$ of Eq.~(\ref{eq:delta}).
Indeed,
\begin{align}
    L(r)
         &= 4\pi^2 R^2 I(R) \frac{R_{\rm ph}^2}{R^2}  \frac{H^4(R)}{H^2(r)H^2(R_{\rm ph})} \nonumber \\
         &= 27 \frac{L(R)}{H^2(r)} \frac{M^2}{R^2} \left(1-\frac{2M}{R}\right)^2.
\end{align}
At $r\to\infty$, the formula reduces to 
\begin{equation}
    L_\infty=L(R)\cdot 27\frac{M^2}{R^2} \left(1-\frac{2M}{R}\right)^2, \ \ \ \mathrm{when}\  R\leq 3M.
    \label{eq:LinfUCO}
\end{equation}
By Eq.~(\ref{eq:delta}), for $R< 3M$ this is
\begin{equation}
    L_\infty=L(R)\,(1-2M/R)\sin^2\delta.
    \label{eq:LinfCO}
\end{equation}
In fact, we can consider this equation to be generally valid for all $R>2M$, if we take $\sin\delta=1$ for $R\geq 3M$. 
Fig. \ref{fig:LR_Linf} shows the ratio $L_{\infty}/L(R)$ of Eq. (\ref{eq:LinfCO}) as a function of the stellar radius. When there is no photon sphere, the profile follows  Eq.~(\ref{eq:LoutCO}), where the luminosity is modified in accordance with the squared redshift of the stellar surface. When the emitting surface lies inside the photon sphere, $R<3M$, the profile changes, as shown in Eq. (\ref{eq:LinfUCO}), to accommodate the additional suppression of the observed luminosity related to the absence of photons from outside the escape cone. It is remarkable that the slope of the observed to emitted luminosity ratio, $l\equiv L_\infty/L(R)$, goes to zero where $l$ vanishes: $\mathrm{d}l/\mathrm{d}R=0$ at $R=2M$.

%%%%%%%%%%%%%%%%%%%%%%%%%%%%%%%%%%%%%%%%%%%%%%%%%%%%%%%%%%%%%%%%%%%%%%%%
\section{Critical Eddington luminosity}
%%%%%%%%%%%%%%%%%%%%%%%%%%%%%%%%%%%%%%%%%%%%%%%%%%%%%%%%%%%%%%%%%%%%%%%%
{We define the critical luminosity as the luminosity for which}
the outward radiation pressure balances gravity.
In {Newtonian gravity, with acceleration $g_\mathrm{N}=GM_*/r^2$,}

{the critical luminosity is given by the well known expression for the Eddington luminosity,} $L_{\rm Edd}=4\pi GM_*c/\kappa$,
where $M_*$ denotes the mass of the star, and  $\kappa$ the (composition dependent) electron scattering opacity. 

In the case of compact stars, general relativistic effects become important. The critical luminosity required to balance gravity in the Schwarzschild metric becomes
\begin{equation}
    L_{\rm net}= L_{\rm Edd}/(1-2M/r)^{1/2},
    \label{eq:Lewin}
\end{equation}
since gravitational acceleration is increased to $g_\mathrm{N}/{H(r)}$.
Combined with Eq.~(\ref{eq:LoutCO}),  this would yield the value for the critical luminosity as seen by a distant observer of
\begin{equation}
    L_\infty =L_{\rm Edd}(1-2M/R)^{1/2}, \ \ \ \mathrm{when}\  R\geq 3M,
    \label{eq:Eddobs}
\end{equation}
familiar from the X-ray burst literature \citep[][]{Lewin1993}.

However, in the case of ultracompact stars ($R<3M$ in the Schwarzschild metric) the presence of the escape cone introduces two competing effects. First, only a fraction of the photons (i.e., $\sin^2\delta$) emitted from the surface escape to infinity.

 Second, the remaining photons fall back onto the star; these returning photons bring back their momentum and energy to the photosphere,
imparting a force in the radially inward direction. By spherical symmetry, the incoming momentum flux carried by these photons is the same as the outgoing momentum of the photons that are outside the escape cone at the point of their emission. However, it takes time for the returning photons to complete their geodesic trajectories. 

%+++++++++++++++++++++++++++++++++++++++++++++++++++++++++++++++++
\begin{figure}
    \centering
    \includegraphics[width=\columnwidth]{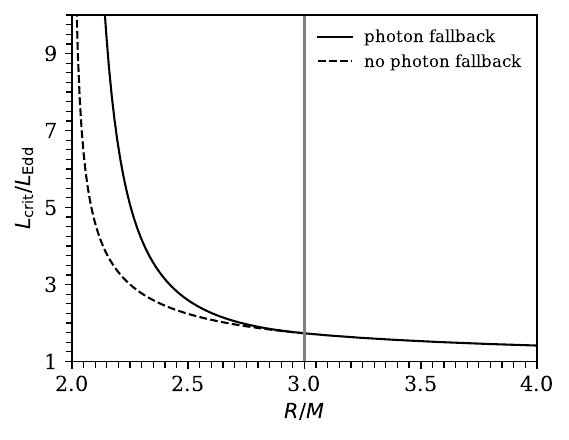}
    \caption{The critical luminosity required to balance gravity is increased owing to the force exerted by the returning photons (fallback) when the stellar surface lies below the photon sphere, located at radius $R=3M$ shown in the gray vertical line.}
    \label{fig:Lcrit_Linf}
\end{figure}
%+++++++++++++++++++++++++++++++++++++++++++++++++++++++++++++++++

We assume a quasi-stationary atmosphere, the requirement being that the radius and other conditions of the photosphere do not vary on the timescale of photon fallback, on the order of $10^2\mu$s for a neutron star, and $10^2\,$s for Sgr A$^*$. Under these assumptions
the net outward flux, and consequently the radiation force, are reduced by the factor  $\sin^2\delta$:
\begin{equation}
    F_\mathrm{net}=F-F(1-\sin^2\delta)=F\sin^2\delta,
    \label{eq:F0}
\end{equation} 
where $F_\mathrm{net}$  represents the net outward flux in the presence of the returning photons,
while the luminosity in outgoing photons is $L(R)=4\pi R^2 F$.
It is the net flux that provides the net radiation force that balances gravity, $F_\mathrm{net}\kappa/c=g_\mathrm{N}/H$, and so the critical value of the luminosity in the outgoing photons $L(R)$
becomes
\begin{equation}
    L_{\rm crit} =  \frac{L_{\rm Edd}}{(1-2M/R)^{1/2}\sin^2\delta}.
    \label{eq:Lcrit}
\end{equation}
Since $\sin^2\delta=1$ for $R\geq 3M$, Eq.~(\ref{eq:Lcrit}) is the general expression for the critical luminosity in the Schwarzschild metric, valid for all $R>2M$.

When the photosphere is within the unstable circular photon orbits ($R<3M$), Eq.~(\ref{eq:Lcrit}) takes the form
\begin{equation}
    L_{\rm crit} = \frac{L_{\rm Edd}R^2}{27M^2(1-2M/R)^{3/2}}.
    \label{eq:Lcrituco}
\end{equation}
The rise in critical luminosity owing to the fall-back of photons is shown in Fig.~\ref{fig:Lcrit_Linf}. The curve rises more steeply within  the photon sphere, compared to the standard formula Eq.~(\ref{eq:Eddobs}).
Clearly, $L_{\rm crit}\rightarrow\infty$ as $R\rightarrow2M$; this reflects how hard it is to resist gravitational collapse in this regime.

On the other hand, since not all the photons escape to infinity, the observed emission $L_\infty=L_{\rm crit}\, H^2\sin^2\delta$ is suppressed by the same factor by which the critical luminosity is enhanced at the photosphere, and finally, for all $R>2M$
\begin{equation}
     L_{\infty}  = L_{\rm Edd}(1-2M/R)^{1/2}.
\label{eq:final}
\end{equation}

This formula for the observed critical luminosity where radiative forces balance gravity at $R\leq 3M$ has the same form as for the standard case of $R>3M$, Eq.~(\ref{eq:Eddobs}). 
%+++++++++++++++++++++++++++++++++++++++++++++++++++++++++++++++++
\begin{figure}
    \centering
    \includegraphics[width=\columnwidth]{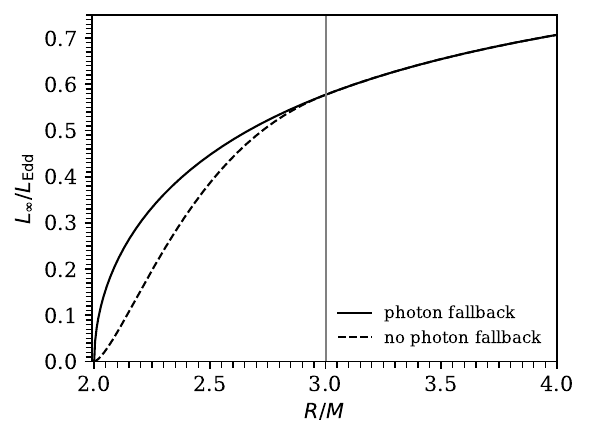}
    \caption{The ratio of the observed emission at critical luminosity to the ordinary Eddington luminosity is plotted as a function of the stellar radius, solid line, Eq.~(\ref{eq:final}).  
    In the X-ray burst literature it has been assumed that the ratio would be suppressed (as shown in the dashed line) owing to the presence of the escape cone when the stellar surface lies below the photon sphere. However, we show that this effect will be compensated by an increase in the critical luminosity required to balance the gravitational force.
    }
    \label{fig:Ledd_Linf}
\end{figure}
%+++++++++++++++++++++++++++++++++++++++++++++++++++++++++++++++++
Previously, in discussion of X-ray bursts the photons outside the escape cone were simply ignored, as though they disappeared below the event horizon, which can only happen in the case of black holes. Thus, it had been thought \cite[e.g.,][]{Lewin1993} that the observed luminosity should be reduced by the factor $\sin^2\delta$ relative to the formula of Eq.~(\ref{eq:final}). This is illustrated in Fig. \ref{fig:Ledd_Linf} for the Schwarzschild case.
The dashed line shows the expected luminosity when the effect of increased critical luminosity owing to photon fallback is ignored. We conclude that
the critical luminosity has been underestimated in the X-ray burst literature, since the inward force from fallback of photons had not been considered.\footnote{Following the completion of this project, we became aware that the critical luminosity, and the observed emission,  for ultracompact objects where the radiative forces balance gravity were already computed correctly by \citet{RL06} in the context of magnetospheric eternally collapsing objects.}

We note that the critical luminosity would be different if spherical symmetry were broken. For instance, if an accretion powered luminosity is released in a narrow equatorial belt (EB), its critical value is given by a formula identical to the one used by \cite{Lewin1993},
\begin{equation}
    L^\mathrm{EB}_\infty\approx L_{\rm Edd}(1-2M/R)^{1/2}\sin^2\delta,
\end{equation}
as the photons outside the escape cone do not necessarily return to the equatorial belt, but mostly to other parts of the stellar surface.

%%%%%%%%%%%%%%%%%%%%%%%%%%%%%%%%%%%%%%%%%%%%%%%%%%%%%%%%%%%%%%%%%%%%%%%%
\section{Discussion}
%%%%%%%%%%%%%%%%%%%%%%%%%%%%%%%%%%%%%%%%%%%%%%%%%%%%%%%%%%%%%%%%%%%%%%%%
%%%%%%%%%%%%%%%%%%%%%%%%%%%%%%%%%%%%%%%%%%%%%%%%%%%%%%%%%%%%%%%%%%%%%%%%
\subsection{Critical luminosity and its minimum observed value }
%%%%%%%%%%%%%%%%%%%%%%%%%%%%%%%%%%%%%%%%%%%%%%%%%%%%%%%%%%%%%%%%%%%%%%%%

\begin{figure}
    \centering
    \includegraphics[width=\columnwidth]{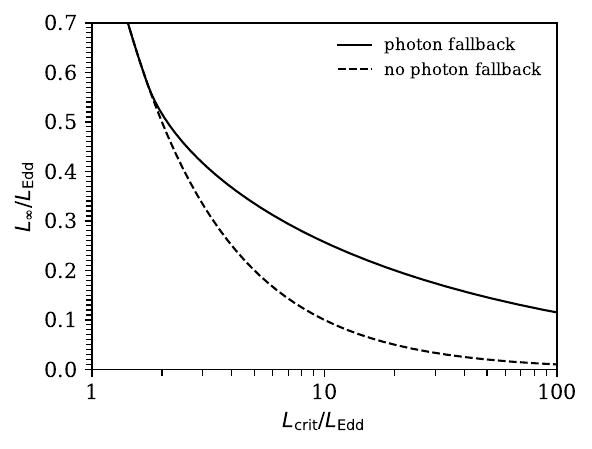}
    \caption{The  
    critical luminosity required to balance the gravitational forces at the photosphere (abscissa) and its value observed at infinity (ordinate).
    }
    \label{fig:Linf_Lcrit}
\end{figure}

 As already noted above, formally,  $L_{\rm crit}\rightarrow\infty$ as $R\rightarrow2M$. However, it is doubtful whether a sufficiently powerful source can be found in practice to provide the requisite critical luminosity at the source for $R\approx 2M$. Presumably, there is an upper limit of the luminosity for any given star, depending on the source of power. For the same reason, even though $L_\infty\rightarrow 0$ as $R\rightarrow2M$, it would seem that if, in fact, the luminosity at the source is critical, there must be a minimum value of $L_\infty$ corresponding to the maximum allowed power $L_{\rm crit}$.
Indeed, we can relate the critical luminosity at the photosphere to its value observed at infinity (Fig.~\ref{fig:Linf_Lcrit})
\begin{equation}
     \left(\frac{L_{\rm crit}}{L_{\rm Edd}}\right) = \left(\frac{L_\infty}{L_{\rm Edd}}\right)^{-1}\frac{1}{\sin^{2}\delta},
\label{eq:fina}
\end{equation}
with $\sin^2\delta=1$ for $R\geq 3M$, as usual, and
\begin{equation}
     \frac{1}{\sin^{2}\delta} = 
     \frac{4}{27}\left(\frac{L_\infty}{L_{\rm Edd}}\right)^{-2}\left[1-\left(\frac{L_\infty}{L_{\rm Edd}}\right)^2\right]^{-2}, \ \ \ \mathrm{when}\  R\leq 3M.
\label{eq:fin}
\end{equation}
It is easy to see from Fig.~\ref{fig:Linf_Lcrit} that if, for instance, a thermonuclear burst has a given power of {\sl at most}  
$xL_{\rm Edd}$, where $x$ is some number, say $x=7$, then (assuming the photospheric radius is sufficiently large for critical luminosity to be reached at some $R$) there is a corresponding {\sl minimum} value of $L_\infty/L_{\rm Edd}$, say, $0.3$.
Paradoxically, the lower the actual value of the critical luminosity at the photosphere, the higher its value observed at infinity. Dimmer bursts are, in fact, more powerful, assuming they all shine at critical luminosity.

%+++++++++++++++++++++++++++++++++++++++++++++++++++++++++++++++++
%+++++++++++++++++++++++++++++++++++++++++++++++++++++++++++++++++

%%%%%%%%%%%%%%%%%%%%%%%%%%%%%%%%%%%%%%%%%%%%%%%%%%%%%%%%%%%%%%%%%%%%%%%%
\subsection{Different nature of the critical luminosity within and without the photon sphere}
%%%%%%%%%%%%%%%%%%%%%%%%%%%%%%%%%%%%%%%%%%%%%%%%%%%%%%%%%%%%%%%%%%%%%%%%
When there is no unstable photon radius ($R>3M$) the critical luminosity, although it has a different (and radius dependent) value in the Schwarzschild metric than the usual Eddington luminosity in Newtonian gravity, reflects the same physical condition---the radiative force acting on the electrons in the atmosphere is balanced by gravity acting on the protons. This gives a local condition on the flux, which can be translated to the luminosity by integrating the flux over a spherical surface. The expression is valid at any instant of time when, and any location where, the local condition of balance is satisfied. 
Whether or not the equilibrium between radiative forces and gravity is stable depends on the local conditions alone. In Newtonian language, the determining factor is whether or not a Lagrangian perturbation (change in the radial position) of the photosphere leads to a change in flux greater or smaller than the change in gravity.

However, for $R<3M$ the condition is a global one on the total luminosity, as the balance between the radiative forces and gravity includes the flux of the returning photons impinging on the photosphere at a particular location but emitted elsewhere. In this regime, as already remarked, Eq.~(\ref{eq:final}) is valid under the assumption of stationarity on the timescale of light travel of the photons between the point of emission and point of return. Stability of equilibrium at any given location now depends on non-local effects, such as the conditions of emission at the origin of the retuning photon's trajectory, and on the presence, or not, of any absorbing, scattering, or emitting matter along the trajectory.
It is quite likely that in this regime the photosphere is prone to spontaneous symmetry breaking, possibly leading to a global instability, e.g. a photospheric expansion to $R>3M$ (a radius expansion in the language of X-ray bursts).

Consider a local "protuberance," whereby a portion of the photosphere forms a column starting to rise above the equilibrium level of the surface. Such an eruption is unstable when $R<3M$.
The top of this column will be subjected to a lower downward flux of momentum imparted by the incoming photons, and so the condition of equilibrium is violated and the column will continue to rise. The rising column will cast a shadow on the nearby areas by intercepting and scattering photons that would have otherwise traveled through the volume now occupied by the column, thus disturbing the equilibrium of nearby areas. At the same time, as the emitting surface rises, the escape cone broadens, and a smaller fraction of photons returns to the photosphere at other locations, so the equilibrium is disturbed even far away from the initial location of the disturbance. On the other hand, a scattering cloud may increase the flux of returning photons. The outcome is hard to predict. Even if a radius expansion of the photosphere does not occur as a result of such instabilities, it seems clear that the spherical symmetry will be broken, so that the photosphere presents a mottled appearance that varies in time, even if the source of luminosity remains uniform on a spherical surface below the photosphere.

It is interesting to note that actual observations of quasiperiodic "burst oscillations" in X-ray bursts suggests a non-uniform photospheric surface long past the expected time of the burning front enveloping the whole surface of the neutron star \citep{B98}.

%%%%%%%%%%%%%%%%%%%%%%%%%%%%%%%%%%%%%%%%%%%%%%%%%%%%%%%%%%%%%%%%%%%%%%%%
\subsection{The observed luminosity when the photospheric luminosity is known}
%%%%%%%%%%%%%%%%%%%%%%%%%%%%%%%%%%%%%%%%%%%%%%%%%%%%%%%%%%%%%%%%%%%%%%%%

The ratio of $L_\infty/L(R)$ can be used to estimate the compactness of a star. 
For ordinary stars $L(R)$ as a function of age and initial mass and metalicity follows from theoretical calculations. Here, we are more interested in stars ultimately powered by accretion, and would like to see if it is possible to determine whether or not the star is within the photon sphere.
%+++++++++++++++++++++++++++++++++++++++++++++++++++++++++++++++++
\begin{figure}
    \centering
    \includegraphics[width=\columnwidth]{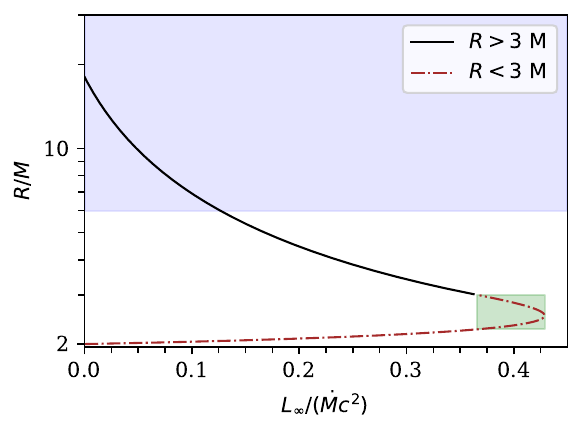}
    \caption{Compactness of the star vs. the observed luminosity,  Eq.~(\ref{eq:KW}), in units of mass accretion rate, $L_{\infty}/(\dot{M}c^2)$, in the case where accretion occurs from the ISCO. The blue shaded region is excluded by the condition that the star is within the ISCO, $R<6M$. For the highest luminosities (green box), $R<3M$, necessarily. For the lowest luminosities, $R<<3M$. For intermediate values of  $L_{\infty}/(\dot{M}c^2)$ two solutions for $R/M$ are possible.}
    \label{fig:RMLMdot}
\end{figure}
%+++++++++++++++++++++++++++++++++++++++++++++++++++++++++++++++++

We begin with the case of luminosity powered directly by accretion. For a given star, and a given accretion configuration (whether thin disk or spherical accretion), the luminosity at the source is now a direct function of $\dot M$, assuming that the accretion rate is low enough for the luminosity to be subcritical. 
However, the observed luminosity is proportional to $\sin^2\delta$. We will illustrate this for spherical accretion. The binding energy released at the stellar surface per unit accreted mass is $H^{-1}-1$, and
\begin{equation}
    L(R)=(H^{-1}-1)H^{-1}\dot M c^2,
\end{equation}
and hence
\begin{equation}
    L_\infty=(1-H)\dot M c^2\sin^2\delta,
    \label{eq:LRH}
\end{equation}
by Eq.~(\ref{eq:LinfCO}), If the accretion proceeds instead through an accretion disk, the term $1-H(r)$ should be replaced by $E_0 - H(R)$, where $E_0$ is the energy per unit mass of the fluid at the inner edge of the disk,
\begin{equation}
    L_\infty=(E_0-H)\dot M c^2\sin^2\delta.
    \label{eq:LRHE0}
\end{equation}
For instance, for a thin disk terminating at the  ISCO, the binding energy released at the stellar surface is the difference between its value in the  marginally stable orbit at $r=6M$ and the surface at $R$, yielding $E_0=\sqrt{8/9}$ \citep{KW85}, and
\begin{equation}
L_\infty=\left(\sqrt{8/9}-\sqrt{1-2M/R}\,\right)\dot M c^2\sin^2\delta(R).
    \label{eq:KW}
\end{equation}
This is an expression for the "instantaneous" emission.
The steady state, bolometric luminosity would be \citep{SS86}
$$L_\infty=\left(\sqrt{8/9}-\sqrt{1-2M/R}\right)\dot M c^2,$$
assuming an albedo of one (reflective inner boundary) and insignificant heat capacity of the accretion-belt fluid, since all the accreted energy must eventually escape the system under these assumptions. We recall that the star under discussion was assumed to be non-rotating.

Under favorable circumstances, Eq.~(\ref{eq:LRHE0}) may be used to rule out a source as a regular, $R>3M$, compact star, particularly if we have an upper limit to its size. For instance the radius of the star could be limited by the inner radius of an accretion disk, as determined via spectral measurements implying the disk temperature profile, and the mass accretion rate can be inferred from the disk luminosity \citep{FRW}. The inner radius of the disk may also be inferred from timing, revealing characteristic frequencies of flux fluctuations \citep{Lyubarskii97,Uttley+01}, or the more prominent quasi-periodic brightness oscillations \citep{VDK00}. 

As an example, the compactness is plotted in Fig.~\ref{fig:RMLMdot} as a function of the ratio $L_\infty/(\dot{M}c^2)$ for a star that accretes from the ISCO, i.e., when it is known that the inner radius of the accretion disk is at $r=6M$. In case of an observation of luminosity corresponding to the green shaded region, it is then in principle possible to directly confirm the presence of an ultracompact ($R<3M$) star, because there are no solutions in that region for a regular compact star. For the rest of the solution domain, e.g., definitely for $L_\infty<0.35\dot Mc^2$, in the absence of independent constraints, the observed luminosity alone does not allow to determine whether or not the star is ultracompact, since the solution is degenerate (there are two branches of the curve in the figure). However, if there is a known upper bound to the stellar radius ($R<6 \rm M$ in  Fig.~\ref{fig:RMLMdot}), and if the observed luminosity is below that predicted by the upper branch of the curve for the upper bound on the radius, say $L_{\infty}<0.1\dot{M}c^2$ in the figure, then the solution for $R/M$ is necessarily on the lower branch of the curve, with $R<<3M$.

%%%%%%%%%%%%%%%%%%%%%%%%%%%%%%%%%%%%%%%%%%%%%%%%%%%%%%%%%%%%%%%%%%%%%%%%
\subsection{Suppressed photon cooling}
%%%%%%%%%%%%%%%%%%%%%%%%%%%%%%%%%%%%%%%%%%%%%%%%%%%%%%%%%%%%%%%%%%%%%%%%
Since the escaping luminosity is reduced by a factor of $\sin^2\delta$, when $R< 3M$ the associated photon cooling time is expected to be increased by $1/\sin^2\delta$.  For X-ray bursts, we would expect the decay of the flux to be slower, and so the total duration of the burst to be longer by a factor of  $1/\sin^2\delta$.
 This can also affect the late-stage thermal cooling \citep{YP04} in very compact stars.

%%%%%%%%%%%%%%%%%%%%%%%%%%%%%%%%%%%%%%%%%%%%%%%%%%%%%%%%%%%%%%%%%%%%%%%%
\section{Non-isotropically emitting stars within the photon sphere}
%%%%%%%%%%%%%%%%%%%%%%%%%%%%%%%%%%%%%%%%%%%%%%%%%%%%%%%%%%%%%%%%%%%%%%%%
We discussed the implications for the observed luminosity of a stellar surface lying within the photon sphere of compact spherical, isotropically emitting stars. 
 
Below we briefly describe some additional implications of a star being so compact that it lies within the photon sphere, when the assumption of isotropy is relaxed.
\begin{enumerate}
 \item In X-ray pulsars the emission may be in a "pencil beam" or a "fan beam," the latter being expected particularly for high luminosities when the polar magnetic flux tube confines the accretion column \citep{GS73,BS75,BS76,Zhang+25}.

 X-ray pulsars observed to emit a fan beam would then likely have a canonical radius $R>3M$ as the fan beams, being directed nearly parallel to the surface, would most likely be outside the escape cone of an $R<3M$ star. Hence, observation of prominent fan beams can rule out the possibility of the source being an ultracompact star. The argument fails if the beam is emitted far above the surface, so that the height of emission added to the stellar radius yields more than $3M$, even when the radius by itself were to satisfy $R<3M$. In pencil beams, the photons are not expected to undergo significant obstruction since they travel along the vertical field lines \citep{S07}, i.e. close to the axis of the escape cone.
 \item
 If the intensity is a function of the polar angle, as in  studies on neutron star atmospheres which show  limb-darkening effects, as shown in \citet{Zavlin+1996},
 limb darkening will be affected by the escape cone when the photosphere is within the unstable circular photon orbits. The narrower the escape cone (as $R/M$ diminishes), the less prominent limb darkening will be, as the lower intensity beams at high polar emission angles will not make it to the observer at infinity.
 This also affects the observed spectrum, since radiation at higher polar angles correspond to lower color temperature, as in Lambert's law.
  \item Pulse profile from hotspots on the surface of neutron star is strongly affected by gravitational bending of light \citep{P+83,Nemiroff+93}, since hotspots in the far side from the observer can become visible too. The modulation is expected to be impacted further in the case of very compact objects where the photon may 
 
 graze the photon sphere, possibly even undergoing several revolutions around the star before departing towards the observer, which would lead to further elongation of the hotspot images \citep{G22}, even if the degree of rotational  modulation would not be reduced significantly \citep{Nemiroff+93}.
\end{enumerate}

\section{Conclusions}
For a fixed value of the photospheric luminosity at the source, the observed luminosity is a monotonically increasing function of the photospheric radius in units of the gravitational radius (Fig.~\ref{fig:LR_Linf}). The critical (gravity-balancing) luminosity is a monotonically decreasing function of the same radius (Fig.~\ref{fig:Lcrit_Linf}). As a result, the observed luminosity is also a monotonically decreasing function of the critical luminosity (Fig.~\ref{fig:Linf_Lcrit}), if the intrinsic luminosity is critical.
If we know the upper limit to the intrinsic luminosity---perhaps following from accretion at a known rate, or considerations of how much thermonuclear fuel can be burned in a unit of time---the observed luminosity of an intrinsically critical luminosity source has a lower limit: intrinsically brighter critical-luminosity sources will be observed as dimmer.

The observed emission is redshifted compared to its value at the surface. In the case of ultracompact stars (of photospheric radius less than that of the unstable circular photon orbit), the observed luminosity undergoes further suppression, since not all of the emitted photons escape to infinity. The suppression factor, $\sin^2\delta$, is related to the escape cone apex half-angle $\delta$.
For the same reason, in X-ray bursts the decay of the flux should be slower, and so the total duration of the burst longer, by a factor of  $1/\sin^2\delta=(1/27)(R/M)^2(1-2M/R)^{-1}$, when $R< 3M$.

The $\sin^2\delta$ suppression of escaping luminosity is not expected to affect the observed critical luminosity (relativistic analogue of Eddington luminosity). The fallback of photons exerts an inward radial force, and hence a higher surface luminosity is required to balance gravity, precisely compensating the effect of the loss cone  \citep{RL06}.
Assuming that the spherically symmetric source shines at critical luminosity  \citep[as in flat-top X-ray bursts,][]{Lewin1993},
the luminosity observed at infinity is a simple function of the radius to mass ratio, $ L_{\infty}  = L_{\rm Edd}(1-2M/R)^{1/2}$, whether or not the photosphere lies within the unstable circular photon orbits.

The above statements apply to isotropically emitting surfaces in the Schwarzschild space-time.
These conclusions need to be interpreted with a few caveats. The outward force on a photosphere from radiation pressure is expected to acquire some modifications related to limb-darkening, e.g., following Lambert's law \citep{Nemiroff+93}. This effect was not considered in this work while deriving the quantities.

Also, the expression for critical luminosity is applicable only for a photosphere in a quasi-static equilibrium. During the onset of an X-ray burst, the timescale of the rapid rise of luminosity
can be comparable to the fall-back time of the photons, and the expression for critical luminosity described in this work may not be applicable in that phase.

We note further that observation of prominent fan beams from hot spots can rule out the possibility of the star being an ultracompact object. Since fan beams are directed parallel to the surface, they are expected to be outside the escape cone when $R<3M$, and unable to escape to infinity.

\section{Acknowledgments}
We thank Dr. Maciek Wielgus and Dr. Leszek Zdunik for discussions.
This work was supported in part by the Polish NCN grant 2019/35/O/ST9/03965.

\bibliographystyle{aa}

\end{document}